\documentclass[11pt,a4paper]{article}
\usepackage{jheppub}

\usepackage{physics}
\usepackage{tensor}
\usepackage{amsmath}
\usepackage{amssymb}
\usepackage[colorlinks=true]{hyperref}
\usepackage{xcolor}
\usepackage{cancel}
\usepackage{ytableau}

\usepackage{titlesec}
\makeatletter
\g@addto@macro\bfseries{\boldmath}
\makeatother

\ytableausetup{boxsize=0.3em,centertableaux}

\usepackage{framed}
\usepackage[most]{tcolorbox}
\usepackage{xcolor}
\colorlet{shadecolor}{gray!10}
\usepackage{marginnote} 

\newcommand{\nn}{\nonumber}

\newcommand{\beq}{\begin{equation}}
\newcommand{\eeq}{\end{equation}}

\preprint{UUITP-21/26} 

\title{(Anti-)De Sitter null strings\\ and Carroll-Weyl symmetry}

\author[a]{Ulf Lindstr\"{o}m}
\author[b]{Bo Sundborg}

\affiliation[a]{Department of Physics and Astronomy, Uppsala University,
Box 516, SE-75120 Uppsala, Sweden
and Centre for Geometry and Physics, Uppsala University,
Box 480, SE-75106 Uppsala, Sweden
}

\emailAdd{ ulf.lindstrom@physics.uu.se}

\affiliation[b]{The Oscar Klein Centre \& Department of Physics, Stockholm University,\\
AlbaNova, 106 91 Stockholm, Sweden}
\emailAdd{bo.sundborg@fysik.su.se}

\abstract{We construct a $dS_d$ null string in a novel way by lifting the standard tensionless string to $R^{1,d}$. Gauging worldsheet scale symmetry introduces terms in the action breaking translation invariance, but preserving the $dS_d$ symmetry $SO(1,d)$. Choosing a De Sitter symmetric gauge fixing of the scale symmetry yields a $d$ dimensional theory which we show has all the expected properties of $dS_d$ tensionless strings. The action happens to be the intriguing Carrol-Weyl symmetric string action in $R^{1,d}$, now with a clear target space interpretation in $d$ dimensions. That it does not, as recently claimed, represent a ($d+1$)-dimensional string in flat ($d+1$)-dimensional Minkowski space settles an apparent issue with the Carrol-Weyl string model. Furthermore, while there are clear obstacles for similar formulations of tensile strings in (A)dS backgrounds, our new insights are easily modified to the formulation of Anti-De Sitter tensionless strings. That our construction generates a curved geometry purely algebraically may have even wider applications.}

\begin{document}
\pagestyle{myplain}
\maketitle

\section{Introduction}
Worldlines of massive particles boosted to infinite momentum approach null worldlines corresponding to massless particles. Similarly, worldsheets of (tensile) strings boosted to infinite momentum density approach null worldsheets of tensionless strings. The subject has a long history with two somewhat disparate beginnings, \cite{Schild:1976vq} and 
\cite{Karlhede:1986wb}. \footnote{Recent development makes it important to discriminate between tensionless strings, which correspond to limits of ordinary strings, and more general null strings which are not necessarily such limits, but where the world sheet is null.} 

Tensionless strings play an interesting role in AdS spacetime and AdS/CFT \cite{Sundborg:1999ue,Sundborg:2000wp}, but strings in dS spacetime and dS holography are shrouded in mystery. It may thus be worthwhile to find new descriptions of null strings in dS spacetime.

In our construction, all steps leading to a curved geometry are purely algebraic.





\section{The De Sitter null string}
Tensionless strings in Minkowski spacetime $M_d$ with metric $\eta_{mn}$ with $m,n=0,1,\dots,d-1$ are described by the action \cite{Isberg:1993av}
\begin{equation}
    S_0=\int d^2\sigma V^aV^b \partial_aX^m\partial_bX^n\eta_{mn} \ .
\end{equation}
which is invariant under the Poincaré group $ISO(1,d-1)$. Our idea for describing De Sitter tensionless strings is based on $d$ dimensional De Sitter space $dS_d$ being a quadric in a $d+1$ dimensional Minkowski space with metric $\eta_{MN}$ with $M,N=0,1,\dots,d$. For physics to reduce to a $d$ dimensional theory, a gauge symmetry is needed, and we introduce a gauge field $W_a$ together with the local gauge transformations
\begin{subequations}\label{eq_ScaleSymmetry}
\begin{align}
  \delta X^M&= \phi(\sigma^a)X^M \label{eq_ScaleSymmetryX}\\ 
\delta V^a&=-\phi(\sigma^a)V^a \label{eq_ScaleSymmetryV}\\
\delta W_a &= -\partial_a\phi(\sigma^a) \label{eq_ScaleSymmetryW}
\end{align}
\end{subequations}
in a gauged action
\begin{equation}
    S=\int d^2\sigma V^aV^b D_aX^MD_bX^N\eta_{MN}
    = \int d^2\sigma V^aV^b (\partial_a+W_a)X^M(\partial_b+W_b)X^N\eta_{MN} \ , \label{eq_GaugedAction}
\end{equation}
with the covariant derivative $D_a=\partial_a +W_a$. We will see explicitly in the Hamiltonian analysis how gauge symmetry reduces the spacetime dimension.\footnote{The gauged action was previously introduced in higher dimensional Dirac space in \cite{Gustafsson:1994kr} where it was called the conformal string.  In \cite{Lindstrom:2026quz,Lindstrom:2026zno} this model was recently shown to reduce to (\ref{eq_GaugedAction}) in lower dimension. This includes constraints and symmetries as well.}  The gauged action \eqref{eq_GaugedAction} was written down earlier this year \cite{Sheikh-Jabbari:2026vqh} and promoted as an action for a null string in flat $(d+1)$-dimensional spacetime, an interpretation which we, however, fail to understand. Gauging of scale symmetry \eqref{eq_ScaleSymmetry} means that scale variation is not physical, which reduces the effective dimension of the target space.

Note that the gauge field terms in the action breaks translation symmetry, reducing the global symmetry $ISO(d,1)$ of the original action to $SO(1,d)$ for the gauged action, which indeed is the isometry group of $dS_d$. We now proceed to demonstrate the reduction of the target space dimension to $d$ by Dirac analysis in the Hamiltonian formulation. Such a reduction should produce De Sitter space as an orbit of $SO(1,d)$. 

\section{Dirac analysis}
The Hamiltonian of this system is
\begin{equation}
    H= \int d\sigma^1\left(\lambda_{P^2}P_M P_N \eta^{MN}+ \lambda_{X'\cdot P} X'^M P_M + \lambda_{X\cdot P} X^M P_M\right)
\end{equation}
with $X'^M\equiv \partial_{\sigma^1}X^M$. As is typical for a reparametrisation invariant system the Hamiltonian is a sum over constraints weighted by Lagrange multipliers. The last term  is special to the present gauged system, while the other terms are the same as for the original tensionless string. This Hamiltonian is obtained from the action \eqref{eq_GaugedAction} using the standard Dirac procedure. Here, the analysis shows that $(V,P_V)$ and $(W,P_W)$ decouple completely and can be disregarded. 
The remaining constraints, 
\begin{equation}
    \varphi_{P^2}=P_M P_N \eta^{MN}=0,\quad \varphi_{X'\cdot P}= X'^M P_M =0,\quad \varphi_{X\cdot P}= X^M P_M=0 \ ,
\end{equation}
are first class.

The essential step is now to fix the additional local scale gauge freedom  without interfering with the familiar tensionless string constraints $\varphi_{P^2}=0$ and $\varphi_{X'\cdot P}=0$. A manifestly $SO(1,d)$ invariant gauge fixing constraint should not be necessary, but is convenient to prove De Sitter invariance. We therefore fix the gauge by imposing
\begin{equation}
    \psi_{dS}=\eta_{MN}X^MX^N-R^2=0\ ,
\end{equation}
which neatly constrains the system to the quadric embedding $dS_d$ with De Sitter radius $R$ in $M_{d+1}$. The value of the radius $R\neq 0$ is not a physical quantity in this model since it is an arbitrary gauge fixing constant.\footnote{Rotating $R\to \pm i R$ leads to null strings in Euclidean $AdS$, ie hyperbolic space, with the same symmetry group $SO(1,d)$.} The Poisson brackets
\begin{equation}\label{eq_GaugeFixing}
    \{\psi_{dS}(\sigma),\varphi_{X\cdot P}(\sigma')\}=2\eta_{MN}X^MX^N \delta(\sigma -\sigma')\approx 2R^2 \delta(\sigma -\sigma') \ ,
\end{equation}
and
\begin{equation}
    \{\psi_{dS}(\sigma),\varphi_{P^2}\}(\sigma')\approx 0, \quad 
    \{\psi_{dS}(\sigma),\varphi_{X'\cdot P}(\sigma')\}= 2\eta_{MN}X'^M(\sigma')X^N(\sigma)\delta(\sigma-\sigma')\approx 0
\end{equation}
imply that $\psi_{dS}=0$ fixes the scale variation and this fixing leaves the standard tensionless string constraints unaffected, respectively. We have used `$\approx$" to refer to weak inequality, ie equality provided the constraints (and gauge fixing conditions) are imposed.

In the Dirac procedure, second class constraints can be eliminated by modifying the Poisson bracket to the Dirac bracket which is constructed to ensure that brackets with the second class constraints vanish. The vanishing is achieved by adding correction terms involving the inverse of the Poisson brackets sandwiched between brackets with the second class constraints. In the present case, there is only one pair of second class constraints, $\varphi_{X\cdot P}$ and $\psi_{dS}$, for each $\sigma$. Their ultra-local Poisson brackets \eqref{eq_GaugeFixing} make the Dirac bracket of the general phase space functions $A$ and $B$ simple: The  inverse of the constraint bracket $\{\psi_{dS}(\sigma'),\varphi_{X\cdot P}(\sigma'')\}$ is represented solely by the inverse $\frac{1}{2R^2}$ of the coefficient of the delta function. We may directly check that the second class constraints are effectively eliminated in the Dirac bracket 
\begin{align}
  \{A,B\}_D=\{A,B\}&\nonumber\\
  -\int d\sigma' d\sigma'' &\left[\{A,\varphi_{X\cdot P}(\sigma')\}\{\psi_{dS}(\sigma'),\varphi_{X\cdot P}(\sigma'')\}^{-1}\{\psi_{dS}(\sigma''),B\}\right.\nonumber\\
  &\left.-\{A,\psi_{dS}(\sigma')\}\{\varphi_{X\cdot P}(\sigma'),\psi_{dS}(\sigma'')\}^{-1}\{\varphi_{X\cdot P}(\sigma''),B\}\right]\ , 
\end{align}
or
\begin{equation}
  \{A,B\}_D= \{A,B\}-\frac{1}{2R^2}\int d\sigma' \left[\,\{A,\varphi_{X\cdot P}(\sigma')\}\{\psi_{dS}(\sigma'),B\} -\{A,\psi_{dS}(\sigma')\}\{\varphi_{X\cdot P}(\sigma'),B\}\right]\ . 
\end{equation}
By our choice of gauge fixing $\psi_{dS}=0$, we have ensured that the Dirac bracket is $dS_d$ invariant.

Note that $X^M(\sigma)$ still behave as coordinate fields in the Dirac brackets
\begin{equation}
 \{X^M(\sigma),X^N(\sigma')\}_D= 0     \ ,   
\end{equation}
since the correction terms to the Poisson bracket vanish. However, $P_M(\sigma)$ are no longer canonically conjugate momentum densities:
\begin{equation}\label{eq_XTranslation}
    \{X^M(\sigma),P_N(\sigma')\}_D= \delta^M_N \delta(\sigma-\sigma')
    -\frac{1}{R^2}X^M X^L \eta_{LN} \delta(\sigma-\sigma')     \ .
\end{equation}
Instead, they generate De Sitter translations.

\section{De Sitter symmetries}
Suppressing the local $\sigma$ dependence, the Dirac bracket algebra of $P_M$ is
\begin{equation}
 \{P_M,P_N\}_D= -\frac{\{P_M,X\cdot P\}\{X^2,P_N\}}{2R^2} -\frac{\{P_M,X^2\}\{X\cdot P,P_N\}}{2R^2} =\frac{1}{R^2}M_{MN} \ .     
\end{equation}
where
\begin{equation}
    M_{MN}\equiv X_M P_N- X_N P_M \ , 
\end{equation}
is a Lorentz generator in $R^{1,d}$ that acts as a De Sitter isometry on the physical configuration space. Here and in the following, indices are raised and lowered by the flat metric $\eta_{MN}$.
\begin{equation}
    \{P_M,M_{KL}\}_D= 
    \{P_M,X_K P_L\}_D- \{P_M,X_L P_K\}_D=\dots =\eta_{MK}P_L-\eta_{ML}P_K \ .
\end{equation}
Finally,
\begin{align}\nn
&\{M_{MN}(\sigma),M_{KL}(\sigma')\}_D\\
&=\left[\eta_{MK}M_{NL}(\sigma)-\eta_{ML}M_{NK}(\sigma)
-\eta_{NK}M_{ML}(\sigma)+\eta_{NL}M_{MK}(\sigma)
\right]\delta(\sigma-\sigma') ,
\end{align}
closing and completing the De Sitter isometry algebra written in terms of generators $P_M$ translating the origin $X^M=0$ and Lorentz transformations $M_{MN}$ preserving the origin. Note that not all the $(d+1) + \frac{d(d+1)}{2}$ generators $P_M$ and $M_{MN}$ are independent. In fact, two different complete sets of generators are given by $M_{MN}$ or by $M_{mn},P_m$ where $m,n=0,1,\dots,d-1$.


\section{Canonical coordinates and a De Sitter metric}

Returning to the issue of canonical momentum densities, they encode the geometry of the target space by keeping track of tangent space. Furthermore, the null constraint $0=P_M \eta^{MN}P_N$ can inform us about the physical metric if $P_M$ is related to a canonical momentum density, which we denote $\Pi_M$. The reason is that the kinetic energy probes the metric of spacetime. In practice, suppose that we can write a linear relation
\begin{equation}\label{eq_PfromPi}
    P_M=E_M{}^N(X)\Pi_N \ .
\end{equation}
Then, the constraint
\begin{equation}\label{eq_MassShell}
    0=P_M\eta^{MN}P_N= E_M{}^K(X)\Pi_K \eta^{MN} E_N{}^L(X)\Pi_L \equiv \Pi_K H^{KL}(X) \Pi_L \ ,
\end{equation}
and the tensor $H^{KL}$ looks like an inverse metric in $d+1$ dimensions. Next, the Dirac bracket will be used to find convenient canonical coordinates on the constraint surface, and the new form of the null constraint can be read off to reveal a De Sitter metric $g_{mn}$ in $d$ dimensions. We have also reduced the phase space by differential geometric methods on the quadric $\psi_{dS}=X^M\eta_{MN}X^N -R^2=0$, but we choose to describe the Dirac bracket procedure to illustrate how curvature is dictated by algebra. The geometric method leads to phase space coordinates directly on the reduced phase space, and the results of the ensuing Poisson brackets agree with the Dirac brackets.

If $\Pi_N$ are conjugate momenta to $X^N$, their conjugacy property
\begin{equation}
  \{X^M(\sigma),\Pi_N(\sigma')\}_D=\delta^M{}_N \,\delta(\sigma-\sigma'),  
\end{equation}
yields
\begin{equation}
    \{X^M(\sigma),P_N(\sigma')\}_D=\{X^M(\sigma),E_N{}^K(X)\Pi_K(\sigma')\}_D
    =E_N{}^K(X)\delta(\sigma-\sigma')\ .
\end{equation}
Then the translation property \eqref{eq_XTranslation} of $P_M$ requires the definition
\begin{equation}
    E_N{}^K(X)\equiv\delta^K_N 
    -\frac{1}{R^2}X^K X_N \ ,
\end{equation}
producing
\begin{align}
    H^{KL}=E_M{}^K(X)\eta^{MN} E_N{}^L(X)=\left(\eta^{NK}-\frac{1}{R^2}X^K X^N\right)\left(\delta^L_N 
    -\frac{1}{R^2}X^L X_N\right)\\
    =\eta^{KL}-\frac{2}{R^2}X^K X^L+\frac{1}{R^4}X^KX^L X^NX_N
\end{align}
which on the constraint surface reduces to
\begin{equation}\label{eq_ConstrainedEmbeddingSpaceMetric}
    H^{KL}\approx \eta^{KL}-\frac{1}{R^2}X^K X^L \ .
\end{equation}
The vector $X_L$ is clearly in the kernel of $H^{KL}$ on the constraint surface, which means that $H^{KL}$ cannot be used as a true inverse metric.
However, the $d$ coordinates $X^m$ on the constraint surface ($X^M$ except $X^d$) are generally independent and $H^{mn}$ can be inverted, giving 
\begin{equation}\label{eq_Metric}
    g_{mn}=(H^{-1})_{mn}=\eta_{mn} + \frac{X_mX_n}{R^2-X^k\eta_{kl}X^l}\ .
\end{equation}
The metric $g_{mn}$ is written with a different symbol from the tensor $H^{MN}$ as it is invertible and neatly gives the metric of $dS_d$ in coordinates $X^m$. Recall that the indices used here are not proper tangent space indices, but are raised and lowered with the ambient metric $\eta_{mn}$. These coordinates are simply $d$ of the coordinates $X^M$ of the embedding space $R^{1,d}$.\footnote{Note that solving the dS constraint is possible in many different ways. For example, Beltrami coordinates correspond to the projection $X^M=(X^m,X^d)\to(X^m,\sqrt{R^2-X^2})$.}
For the reduced configuration space after fixing scale symmetry, we have simply chosen $d$ coordinates projected from the embedding space $R^{1,d}$, but the non-linear $\psi_{dS}$ condition forces correction terms to the corresponding conjugate momenta, which leads to the non-trivial metric.

Inverting the relation \eqref{eq_PfromPi} yields the momentum
\begin{equation}
    \Pi_n=P_n+\frac{X_nX^m}{R^2-X^k\eta_{kl}X^l}P_m \ .
\end{equation}
Now, the constraint $0=\varphi_{X\cdot P}=X^mP_m +X^dP_d$ implies
\begin{equation}\label{eq_ProjectedMomenta}
    \Pi_n= P_n-\frac{X_nX^d}{R^2-X^k\eta_{kl}X^l}P_d
    \approx P_n-\frac{X_n}{X^d}P_d
\end{equation}
which confirms that $X^m$ and $\Pi_n$ are conjugate:
\begin{equation}
    \{X^m,\Pi_n\}_D=\{X^m,P_n\}_D-\frac{X_n}{X^d}\{X^m,P_d\}_D=\delta^m_n 
    -\frac{1}{R^2}X^m X_n + \frac{1}{R^2}\frac{X_n}{X^d}X^mX_d =\delta^m_n \ .
\end{equation}
Similarly, but with some more steps,
\begin{align}
    \{\Pi_m, \Pi_n\}_D =& \{P_m-\frac{X_m}{X^d} P_d, P_n-\frac{X_n}{X^d} P_d\}_D=\dots=0 
\end{align}
completes the demonstration that $X^m$ and $\Pi_n$ are canonical coordinates for the Dirac bracket. Finally, we note that the momenta \eqref{eq_ProjectedMomenta} imply that the null constraint \eqref{eq_MassShell}
\begin{equation}
0=P_M\eta^{MN}P_N=\Pi_mH^{mn} \Pi_n\     
\end{equation}
verifies that the physical metric is $g_{mn}=(H^{-1})_{mn}$.

To summarise, the $SO(1,d)$ symmetric gauge fixing has given a spacetime with a non-trivial metric, disclosed by the Dirac bracket, which is the physical Poisson bracket on the gauge fixed phase space. 
One coordinate and its conjugate momentum can be removed from the description, and a simple choice is $(X^m,\Pi_m)$ with $m=0, 1, \dots, d-1$ as canonical variables
.

Finally, Legendre transforming the Hamiltonian of the reduced ($d$-dimensional) model yields a standard tensionless string action
\begin{equation}
    S_0=\int d^2\sigma V^aV^b \partial_aX^m\partial_bX^n g_{mn}(X) \ .
\end{equation}
but now in De Sitter space. This reduced action no longer enjoys the scaling, or Carrol-Weyl symmetry, \eqref{eq_ScaleSymmetry} which was gauge fixed.


\section{Conclusion}
We have found a construction of tensionless string theory in De Sitter space, as defined by an algebraic equation, without explicitly introducing a De Sitter metric. This procedure can be immediately taken over to anti-De Sitter space by starting from $Y^M\in R^{2,d-1}$ instead of  $X^M\in R^{1,d}$ and fixing the gauge $\psi_{AdS}=\eta^{2,d-1}_{MN} Y^MY^N+R^2=0$ in analogy to $\psi_{dS}=0$. Interestingly, the procedure does not work for standard tensile strings with the standard Brink-Di Vecchia-Howe-Polyakov action because its Weyl invariance makes it impossible for a rescaling of the world sheet metric to compensate for a rescaling of the target space. In contrast, the Carrollian worldsheet geometry of null strings allows an additional Carroll-Weyl symmetry, which simultaneously rescales the target space. This symmetry reduces the number of degrees of freedom from the conventional string value of $(d+1)-2$ to $(d+1)-3$ \cite{Sheikh-Jabbari:2026tpf}, suggesting an interpretation as a null string with specially reduced spectrum in $(d+1)$ dimensions.
In this work, we found a more natural interpretation: that the null string propagates in $d$ dimensions where the number of degrees of freedom of the string equals $d-2$. Moreover, the geometry is actually $dS_d$.

There are already signs in the action \eqref{eq_GaugedAction} in $R^{1,d}$ casting doubt on the flat $(d+1)$-dimensional picture, since the original translation invariance is broken in the Carroll-Weyl symmetric gauged action. This means that the coordinates of $R^{1,d}$ are not coordinates of physical Minkowski space $M^{d+1}$, something which also follows from the gauge non-invariance of the coordinates displayed in eq.\ \eqref{eq_ScaleSymmetry}.

The identification of the De Sitter target space may also explain the otherwise surprising discrete spectrum of a recent quantisation procedure \cite{Rasulian:2026jvg}. Quantisation in $dS$ is non-trivial, but this intriguing result would be consistent with the compact spherical spacelike sections of the De Sitter spacetime in global coordinates.

Our novel construction of strings in backgrounds with non-zero cosmological constant opens several avenues of research. Mathematically, other algebraically embedded surfaces than $dS$ or $AdS$ can be contemplated. With $AdS$ strings, questions about the $AdS$ boundary and $AdS/CFT$ can be addressed naturally. It is especially intriguing that precisely the \emph{tensionless} $AdS$ strings, which arise naturally in the holographic dual of $\mathcal{N}=4$ super-Yang-Mills as a bridge to perturbative quantum field theory \cite{Haggi-Mani:2000dxu,Sundborg:2000wp, Minahan:2002ve}, are described by the proposed method. Do generalisations of the present action principle give a fundamentally new perspective on $AdS/CFT$? Even more speculatively, can the $dS$ tensionless strings shed light on holography for quantum gravity in De Sitter space? 

More concretely, De Sitter tensionless strings are interesting in their own right, and reinterpreting recent results on Carroll-Weyl symmetric strings and their quantisation in De Sitter terms will be instructive from both a string theory and a De Sitter perspective.

\acknowledgments
Correspondence with Ida Rasulian, Shahin Sheik-Jabbari and Hossein Yavartanoo is gratefully acknowledged.


\appendix

\bibliographystyle{JHEP}
\bibliography{references}

\end{document}